\documentstyle[aps,pra,multicol]{revtex}
\begin{document}

\draft
\title{The Dilute Bose Gas Revised}
\author{A. Yu. Cherny$^{1}$ and A. A. Shanenko$^{2}$ }
\address{$^{1}$Frank Laboratory of Neutron Physics,
Joint Institute for Nuclear Research,
141980, Dubna, Moscow region, Russia}
\address{$^{2}$Bogoliubov Laboratory of Theoretical Physics,
Joint Institute for Nuclear Research,
141980, Dubna, Moscow region, Russia}
\date{February 17, 1998}

\maketitle

\begin{abstract}
The well--known results concerning a dilute Bose gas with the
short--range repulsive interaction should be reconsidered due to a
thermodynamic inconsistency of the method being basic to much of the
present understanding of this subject.  The aim of our paper is to
propose a new way of treating the dilute Bose gas with an arbitrary
strong interaction.  Using the reduced density matrix of the second
order and a variational procedure, this way allows us to escape the
inconsistency mentioned and operate with singular potentials of the
Lennard--Jones type. The derived expansion of the condensate
depletion in powers of the boson density $n=N/V$ reproduces the
familiar result, while the expansion for the mean energy per
particle is of the new form:$\;\varepsilon=2\pi \hbar^2 a n/m
\{1+128/(15 \sqrt{\pi})\sqrt{na^3}(1-5b/8a)+ \cdots\}$, where $a$
is the scattering length and $b\geq0$ stands for one more
characteristic length depending on the shape of the interaction
potential (in particular, for the hard spheres $a=b$).
All the consideration concerns the zero
temperature.

\end{abstract}

\begin{multicols}{2}

It is well--known that to investigate a dilute Bose gas of particles
with an arbitrary strong repulsion (the strong--coupling regime),
one should go beyond the Bogoliubov approach~\cite{Bog1}
(weak--coupling case) and treat the short--range boson correlations
in a more accurate way. An ordinary manner of doing so is the use of
the Bogoliubov model with the ``dressed", or effective, interaction
potential containing ``information" on the short--range boson
correlations (see Ref. \cite{Lee}). Below it is
demonstrated that this manner leads to a loss of the thermodynamic
consistency. To overcome this trouble, we propose a new way of
investigating the strong--coupling regime which concerns the reduced
density matrix of the second order~(the 2--matrix) and is based on
the variational method.

The 2--matrix for the many--body system of spinless bosons can be
represented as~\cite{Bog2}: $
\rho_2({\bf r}_1^{\prime},{\bf r}_2^{\prime};{\bf r}_1,{\bf r}_2)
= F_2({\bf r}_1,{\bf r}_2;{\bf r}_1^{\prime},{\bf r}_2^{\prime})
/\{N(N-1)\},$
where the pair correlation function is given by
\begin{equation}
F_2({\bf r}_1,{\bf r}_2;{\bf r}_1^{\prime},{\bf r}_2^{\prime})=
\langle \psi^{\dagger}({\bf r}_1) \psi^{\dagger}({\bf r}_2)
    \psi ({\bf r}_2^{\prime})\psi ({\bf r}_1^{\prime})\rangle.
\label{2}
\end{equation}
Here $\psi ({\bf r})$ and $\psi^{\dagger}({\bf r})$ denote the boson
field operators. Recently it has been found~\cite{Chern1,Chern2}
that for the uniform system with a small depletion of the
zero--momentum state the correlation function (\ref{2}) can be
written in the thermodynamic limit as follows~\cite{Chern1,Chern2}:
\begin{eqnarray}
&&F_2({\bf r}_1,{\bf r}_2;{\bf r}_1^{\prime},{\bf r}_2^{\prime})=
n_0^2\,\varphi^*(r)\,\varphi(r^{\prime})
\nonumber\\
&&+2n_0\int \frac{d^3q}{(2\pi)^3}\,
       n_q\,\varphi_{{\bf q}/2}^*({\bf r})
            \varphi_{{\bf q}/2}({\bf r}^{\prime})
                  \exp\{ i {\bf q}({\bf R}^{\prime}-{\bf R})\},
\label{3}
\end{eqnarray}
where ${\bf r}={\bf r}_1 - {\bf r}_2, \;{\bf R}=({\bf r}_1
+{\bf r}_2)/2$
and similar relations take place for ${\bf r}^{\prime}$ and ${\bf
R}^{\prime}$, respectively. In Eq.~(\ref{3}) $n_0=N_0/V$ is the
density of the particles in the zero--momentum state, $n_q=\langle
a_{\bf q}^{\dagger}a_{\bf q}\rangle$ stands for the distribution of
the noncondensed bosons over momenta. Besides, $\varphi(r)$ is the
wave function of a pair of particles being both condensed. In turn,
$\varphi_{{\bf q}/2}({\bf r})$ denotes the wave function of the
relative motion in a pair of bosons with the total momentum
$\hbar{\bf q}$, this pair including one condensed and one
noncondensed particles.
So, Eq.~(\ref{3}) takes into account the condensate--condensate and
supracondensate--condensate pair states and is related to the
situation of a small depletion of the zero--momentum one--boson
state. For the wave functions $\varphi(r)$ and $\varphi_{{\bf p}}
({\bf r})$ we have
\begin{eqnarray}
&&\varphi(r)\!=\!1+\psi(r),\,\varphi_{{\bf p}}\!({\bf r})\!=\!
\sqrt{2}\cos({\bf p}{\bf r})+\psi_{{\bf p}}({\bf r})\;(p\not=0)
\label{4}
\end{eqnarray}
with the boundary conditions $ \psi(r) \to 0$ and
$\psi_{{\bf p}}({\bf r}) \to 0$ for $r \to \infty.$
The functions $\psi(r)$ and $\psi_{{\bf p}}({\bf r})$ can
explicitly be expressed in terms of the Bose operators $a_{{\bf
p}}^{\dagger}$ and $a_{{\bf p}}$ \cite{Chern1}. In particular,
\begin{equation}
\widetilde{\psi}(k)=\int\psi(r)
 \exp(-i{\bf k}{\bf r})\,d^3r=
  \langle a_{{\bf k}}\,a_{-{\bf k}}\rangle/n_0\,.
\label{6}
\end{equation}
Having in our disposal the distribution function $n_{k}$ and the set
of the pair wave functions $\varphi(r)$ and $\varphi_{{\bf p}} ({\bf
r})$, we are able to calculate the main thermodynamic quantities of
the system of interest. In particular, the mean energy per particle is
expressed in terms of $n_k$ and $g(r)$ via the well--known formula
\begin{eqnarray}
\varepsilon=\int \frac{d^3k}{(2\pi)^3} T_k \frac{n_k}{n}+
\frac{n}{2}\int g(r) \Phi(r) d^3r,
\label{7}
\end{eqnarray}
where $T_k=\hbar^2 k^2/2m$ is the one--particle kinetic energy,
$n=N/V$ stands for the boson density and the relation
\begin{equation}
g(r)=F_2({\bf r}_1,{\bf r}_2;{\bf r}_1,{\bf r}_2)/n^2.
\label{8}
\end{equation}
is valid for the pair distribution function $g(r)$.

The starting point of our investigation is the weak--coupling regime
which implies weak spatial correlations of particles and, thus, is
characterized by the set of the inequalities
\begin{equation}
|\psi(r)| \ll 1, \;\quad |\psi_{{\bf p}}({\bf r})| \ll 1\;.
\label{9}
\end{equation}
Specifically, the Bogoliubov model corresponds to the
choice~\cite{Chern1,Chern2}
\begin{equation}
|\psi(r)| \ll 1, \;\quad \psi_{{\bf p}}({\bf r}) = 0\;.
\label{10}
\end{equation}
Besides, owing to a small depletion of the Bose condensate
$(n-n_0)/n$ we have for the one--particle density matrix
$F_{1}(r)=\langle\psi^{\dagger}({\bf r}_{1})\psi({\bf
r}_{2})\rangle$:
$$
\left|\frac{F_{1}(r)}{n}\right|=
\left|\int\frac{d^3k}{(2\pi)^3}\frac{n_k}{n}
\exp(i{\bf k}{\bf r})\right|\leq \frac{n-n_{0}}{n}\ll 1.
$$
So, investigating the Bose gas within the Bogoliubov scheme, we have
two small quantities: $\psi(r)$ and $F_{1}(r)/n$.  This enables us
to write Eq.~(\ref{8}) with the help of (\ref{3}) as follows:
\begin{equation}
g(r)=1+2\psi(r)+\frac{2}{n}
\int\frac{d^3k}{(2\pi)^3}n_k\exp(i{\bf k}{\bf r}),
\label{11}
\end{equation}
where we restricted ourselves to the terms linear in $\psi(r)$ and
$F_{1}(r)/n$ and put $\psi^*(r)=\psi(r)$ because the pair wave
functions can be chosen as real quantities.  Equations for
$\widetilde{\psi}(k)$ and $n_k$ can be found varying the mean energy
(\ref{7}) with (\ref{11}) taken into account. However, previously
one should realize an important point, namely: $n_k$ and
$\widetilde{\psi}(k)$ can not be independent variables.  Indeed,
when there is no interaction between particles, there are no spatial
particle correlations either. So, $\widetilde{\psi}(k)=0$ and, since
the zero--temperature case is considered, all the bosons are
condensed, $n_k=0$. While ``switching on" the interaction results
in appearing the spatial correlations and condensate depletion:
$\widetilde{\psi}(k)\not=0$ together with $n_k\not=0$.  In the
framework of the Bogoliubov scheme $\widetilde{\psi}(k)$ is related
to $n_k$ by the expression
\begin{equation}
n_k (n_k+1)=n_0^2\widetilde{\psi}^2(k).
\label{12}
\end{equation}
Indeed, the canonical Bogoliubov transformation~\cite{Bog1}
implies that
\begin{equation}
a_{{\bf k}}=u_{k}\alpha_{{\bf k}}+v_{k}
\alpha^{\dagger}_{-{\bf k}},\quad
a^{\dagger}_{{\bf k}}=u_{k}\alpha^{\dagger}_{{\bf k}}+
v_{k}\alpha_{-{\bf k}},
\label{12a}
\end{equation}
where
\begin{equation}
u_{k}^2-v_{k}^2=1.
\label{13}
\end{equation}
At zero temperature $\langle\alpha^{\dagger}_{{\bf k}}
\alpha_{{\bf k}} \rangle=0$ and, using Eqs.~(\ref{6}) and (\ref{12a})
we arrive at
\begin{equation}
n_k=v_{k}^2, \quad
      \widetilde{\psi}(k)=u_{k}v_{k}/n_0.
\label{14}
\end{equation}
With Eqs.~(\ref{13}) and (\ref{14}) one can readily obtain
(\ref{12}).

Now, let us show that all the results on the thermodynamics of a
weak--coupling Bose gas can be derived for the Bogoliubov scheme with
variation of the mean energy (\ref{7}) under the conditions
(\ref{11}) and (\ref{12}). Inserting (\ref{11}) into (\ref{7}) and,
then, varying the obtained expression, we arrive at
\begin{equation}
\delta\varepsilon=\int \frac{d^3k}{(2\pi)^3} \left[\Bigl(T_k+
n\widetilde\Phi(k)\Bigr)\frac{\delta n_k}{n} + n\widetilde\Phi(k)\delta
\widetilde\psi(k)\right].
\label{15}
\end{equation}
Relation (\ref{12}) connecting $\widetilde\psi(k)$ with $n_k$
results in
\begin{equation}
\delta\widetilde{\psi}(k)=\frac{(2n_k+1)\delta n_k}{2n_0^2
\widetilde{\psi}(k)}+
\frac{\widetilde{\psi}(k)}{n_0}\int\frac{d^3q}{(2\pi)^3}\delta n_q,
\label{16}
\end{equation}
where the equality
\begin{equation}
n=n_0+\int\frac{d^3k}{(2\pi)^3}\;n_k
\label{17}
\end{equation}
is taken into consideration.
Setting $\delta\varepsilon=0$ and using Eqs.~(\ref{15}) and
(\ref{16}), we derive the following expression:
\begin{eqnarray}
-2\,T_k &&\widetilde{\psi}(k)=
     \frac{n^2}{n^2_0} \widetilde{\Phi}(k)(1+2n_k)
        \nonumber\\
&&+2n\,\widetilde{\psi}(k)\left(\widetilde{\Phi}(k)+\frac{n}{n_0}
     \int\frac{d^3q}{(2\pi)^3}
           \widetilde{\Phi}(q)\widetilde{\psi}(q)\right).
\label{18}
\end{eqnarray}
Here one should realize that Eq.~(\ref{18}) is able to yield results
being accurate only to the leading order in $(n-n_0)/n$ because the
used expression for $g(r)$ given by (\ref{11}) is valid to the
next--to--leading order~\cite{note4}. So, Eq.~(\ref{18}) should be
rewritten as
\begin{equation}
-2\,T_k \widetilde{\psi}(k)=
\widetilde{\Phi}(k)(1+2n_k)+2n\,\widetilde{\psi}(k)\Phi(k).
\label{19}
\end{equation}
Eq.~(\ref{19}) is an equation of the Bethe--Goldstone type or, in
other words, the in--medium Schr\"odinger equation for the pair wave
function. As $2\widetilde{\Phi}(k)(n_k+n \widetilde{\psi}(k))$ is
the product of the Fourier transforms of $\Phi(r)$ and $n(g(r)-1)$,
we can rewrite Eq.~(\ref{19}) in the more customary form
\begin{equation}
\frac{\hbar^2}{m}\nabla^2\varphi(r)=\Phi(r)
+n\int\Phi(|{\bf r}-{\bf y}|)\Bigl( g(y)-1\Bigr)d^3y.
\label{20}
\end{equation}
The structure of Eq.~(\ref{20}) is discussed in the
papers~\cite{Chern2,Shan1}. Here we only remark that the right--hand
side (r.h.s.) of (\ref{20}) is the in--medium potential of the
boson--boson interaction in the weak--coupling approximation. The system of
equations (\ref{12}) and (\ref{19}) can easily be solved, which leads
to the familiar results~\cite{Bog1}:
\vspace*{-2mm}
\begin{eqnarray}
&&n_k=\frac{1}{2}\left(\frac{T_k+n\widetilde{\Phi}(k)}{\sqrt{T_k^2+
2nT_k\widetilde{\Phi}(k)}}-1\right),\nonumber\\[-1mm]
&&\widetilde{\psi}(k)=-\frac{
\widetilde{\Phi}(k)}{2\sqrt{T_k^2+2nT_k\widetilde{\Phi}(k)}}\,.
\label{21}
\end{eqnarray}

Now we are able to demonstrate that the investigation of the
strong--coupling case based on the Bogoliubov model with the
effective boson--boson interaction, results in a loss of the
thermodynamic consistency. Indeed, as it was shown in the previous
paragraph, any calculating scheme using the basic relations of the
Bogoliubov model (\ref{11}), (\ref{12}) conclusively leads to
Eqs.~(\ref{19})-(\ref{21}) provided this scheme does yield the
minimum of the mean energy.  In this case Eqs.~(\ref{19})-(\ref{21})
certainly includes the quantity $\Phi(r)$ which is the ``bare"
interaction potential appearing in (\ref{7}). The use of the
Bogoliubov model with the effective interaction potential
substituted for $\Phi(r)$ can in no way disturb the relations given
by (\ref{11}) and (\ref{12}). And Eq.~(\ref{7}) is the same in both the
weak-- and strong--coupling regimes. Thus, any attempts of replacing
$\Phi(r)$ by the effective ``dressed" potential without
modifications of (\ref{11}) and (\ref{12}) results in a calculating
procedure which does not really provide the minimum of the mean
energy. It is nothing else but a loss of the thermodynamic
consistency. Remark that we do not mean, of course, that the
t--matrix approach or the pseudopotential method can not be applied
in the quantum scattering problem. It is only stated that the usual
way of combining the ladder diagrams with the random phase
approximation faces the trouble mentioned above. Though our
present investigation is limited by the consideration of the Bose
systems, the derived result gives a hint that the similar situation is
likely to take place in the Fermi case, too. In this connection it
is worth noting the problem associated with the lack of
self--consistency of the standard method of treating the dilute
Fermi gas~\cite{Fetter}.

The strong--coupling regime is characterized by significant spatial
correlations. So, Eq.~(\ref{10}) resulting in (\ref{11}) is not
relevant for an arbitrary strong repulsion between bosons at small
separations when we have $\psi(0)=-1, \quad \psi_{{\bf p}}(0)=
-\sqrt{2}$ (see Refs.~\cite{Chern1,Chern2}).
Therefore, to investigate the strong--coupling regime,
Eq.~(\ref{11}) should be abandoned in favor of (\ref{3}).
Expression (\ref{3}) is accurate to the next--to--leading order in
$(n-n_0)/n$. So, using (\ref{3}) and (\ref{8}), we can write
\begin{equation}
g(r)=\varphi^2(r)+\frac{2}{n}\int\frac{d^3q}{(2\pi)^3} n_q
\left(\varphi^2_{{\bf q }/2}({\bf r})-\varphi^2(r) \right).
\label{25}
\end{equation}
Let us now perturb $\widetilde{\psi}(k)$ and $n(k)$. Working to
the first order in the perturbation and keeping in mind conditions
(\ref{12}) and (\ref{25}), from (\ref{7}) we derive:
\begin{equation}
-2\,T_k \widetilde{\psi}(k)=
    \widetilde{U}(k)(1+2n_k)+2n\,\widetilde{\psi}(k)
                               \widetilde{U}^{\prime}(k)
\label{26}
\end{equation}
with
\begin{equation}
\widetilde{U}(k)=\int \varphi(r) \Phi(r) \exp(-i{\bf k}{\bf r})
d^3r
\label{27}
\end{equation}
and
\begin{eqnarray}
\widetilde{U}^{\prime}(k)=
 \int\Bigl(\varphi_{{\bf k}/2}^2({\bf r})
   -\varphi^2(r)\Bigr)\,\Phi(r)\,d^3r.
\label{28}
\end{eqnarray}
Using Eqs.~(\ref{27}), (\ref{28}) as well
as the relation $\psi_{\bf k}({\bf r}) \to \sqrt{2}\psi(r)\;(k \to
0)$ (see the boundary conditions (\ref{4})), we obtain
$\widetilde{U}(0)\not=\widetilde{U}^{\prime}(0).$ This implies that
the system of Eqs.~(\ref{12}) and (\ref{26}) is not able to yield
the relation $n_k\propto 1/k\;(k \to 0)$ following from the
``$1/k^2$" theorem of Bogoliubov for the zero
temperature~\cite{Bog3}. Indeed, let us assume $n_k \to \infty$ for
$k \to 0.$ Then, from Eq.~(\ref{12}) at $n=n_0$ we find
$n|\widetilde{\psi}(k)|/n_k \to 1$ when $k \to 0.$ On the contrary,
Eq.~(\ref{26}) gives $n|\widetilde{\psi}(k)|/n_k \to
\widetilde{U}(0)/\widetilde{U}^{\prime}(0)\not=1$ for $k \to 0.$
So, consideration of the Bose gas based on Eqs.~(\ref{3}) and
(\ref{12}) does not produce satisfactory results. Nevertheless, it
is worth noting that Eq.~(\ref{26}) has an important peculiarity
which differentiate it from Eq.~(\ref{19}) in an advantageous way.
The point is that in both the limits $n\to 0$ and $k \to \infty$
Eq.~(\ref{26}) is reduced to
\begin{equation}
-\frac{\hbar^2}{m}\,\nabla^2\,\varphi(r)+\Phi(r)\varphi(r)=0.
\label{28aa}
\end{equation}
As it is seen, this is the exact ``bare" (not
in--medium) Schr\"odinger equation, other than its
Born approximation following from (\ref{20}). Thus, we can expect
the line of our investigation to be right.

As it was shown in the previous paragraph, an approach adequate for
a dilute Bose gas with an arbitrary strong interaction can not be
constructed without modifications of Eq.~(\ref{12}). This is also in
agreement with a consequence of the relation
\begin{equation}
|\langle a_{{\bf k}}\,a_{-{\bf k}}\rangle|^2 \leq
\langle a_{{\bf k}}\,a_{{\bf k}}^{\dagger}\rangle
\langle a_{-{\bf k}}^{\dagger}\,a_{-{\bf k}}\rangle
\label{29}
\end{equation}
resulting from the inequality of
Cauchy--Schwarz--Bogoliubov~\cite{Bog3}
$
|\langle \widehat{A}\widehat{B}\rangle|^{2} \leq
    \langle\widehat{A}\widehat{A}^{\dagger}\rangle
    \langle\widehat{B}^{\dagger}\widehat{B}\rangle.
$
With (\ref{6}) and (\ref{29}) one can easily derive $n_0^2
\widetilde{\psi}^2(k) \leq n_k(n_k+1)$. Thus, it is reasonable
to assume that Eq.~(\ref{12}) takes into account only the
condensate--condensate channel and ignores the
supracondensate--condensate ones. Now the question arises
how to find corrections to the r.h.s. of Eq.~(\ref{12}).
At present we have no regular procedure allowing us to do this
in any order of $(n-n_0)/n$. However, there exists an argument
which makes it possible to realize the first step in this
direction. The matter is that the alterations needed have to
produce the equation for $\widetilde{\psi}_{\bf p}({\bf k})$
which is reduced to the equation for $\widetilde{\psi}(k)$
in the limit $p \to 0.$ Though this requirement does not
uniquely determine the corrections to Eq.~(\ref{12}), it
turns out to be significantly restrictive. In particular,
even the simplest variant of correcting Eq.~(\ref{12}) in this
way, leads to promising results. Indeed, this variant is specified
by the expression
\begin{eqnarray}
n_k(n_k+1)= n_0^2\,\widetilde{\psi}^2(k)
+2 n_0\int \frac{d^3q}{(2\pi)^3}\,n_q
                   \widetilde{\psi}^2_{{\bf q}/2}({\bf k}).
\label{30}
\end{eqnarray}
Eq.~(\ref{30}) is valid to the next--to--leading order in
$(n-n_0)/n$. So, we may rewrite it as
\begin{equation}
n_k(n_k+1)=n^2\widetilde{\psi}^2\!(k)+2n\!\int\!
\frac{d^3q}{(2\pi)^3} n_q\!\left(\!\widetilde{\psi}^2_{{\bf q }/2}
({\bf k})-\widetilde{\psi}^2(k)\!\right)\!.
\label{31}
\end{equation}
Perturbing $\widetilde{\psi}(k)$ and $n_k$ and bearing in mind
conditions (\ref{25}) and (\ref{31}), (\ref{7}) gives
Eq.~(\ref{26}) again. However, now $\widetilde{U}^{\prime}(k)$
obeys the new relation
\begin{eqnarray}
\widetilde{U}^{\prime}(k)&=&
 \int\Bigl(\varphi_{{\bf k}/2}^2({\bf r})
         -\varphi^2(r)\Bigr)\,\Phi(r)\,d^3r\nonumber\\
&&-\int\frac{d^3q}{(2\pi)^3}
  \frac{\widetilde{U}(q)
    \bigl(\widetilde{\psi}^2_{{\bf k}/2}({\bf q})-
       \widetilde{\psi}^2(q)\bigr)}{\widetilde{\psi}(q)}
\label{33}
\end{eqnarray}
which significantly differs from (\ref{28}). Indeed,
the choice of the pair wave functions as real quantities
implies that operating with integrands in (\ref{27}) and
(\ref{33}), one can exploit $\psi_{{\bf p}}({\bf r})-\sqrt{2}
\psi(r) \propto p^2$ at small $p$~\cite{note3}. For $k \to 0$
this provides $\widetilde{U}^{\prime}(k)-\widetilde{U}(k)=
t_k= c\,k^4 +\cdots.$ Similar to Eq.~(\ref{19}), Eq.~(\ref{26})
can yields results correct only to the leading order in
$(n-n_0)/n$. So, it has to be solved together with (\ref{12})
where $n_0^2$ should be replaced by $n^2$, rather than with
(\ref{31}). This leads to the following relation:
\vspace*{-2mm}
\begin{eqnarray}
&&n_k=\frac{1}{2}\left(\frac{\widetilde{T}_k+n\widetilde{U}(k)}
{\sqrt{\widetilde{T}_k^2+2n\widetilde{T}_k\widetilde{U}(k)}}
-1\right),\label{35a}\\[-1mm]
&&\widetilde{\psi}(k)=-\frac{
\widetilde{U}(k)}{2\sqrt{\widetilde{T}_k^2+2n\widetilde{T}_k
\widetilde{U}(k)}},
\label{36}
\end{eqnarray}
where $\widetilde{T}_k=T_k+nt_k$. In the limit $k \to 0$
Eq.~(\ref{36}) gives $n_k \simeq (\sqrt{n\,m\,\widetilde{U}(0)}/
\hbar k-1)/2$, which is fully consistent with the ``$1/k^2$"
theorem of Bogoliubov for the zero temperature~\cite{Bog3}.
Eqs.~(\ref{27}) and (\ref{36}) should be solved in a
self--consistent manner. So, for $n \to 0$ one can derive
\begin{equation}
\widetilde{U}(k)=\widetilde{U}^{(0)}(k)(1+8\sqrt{na^3}/
\sqrt{\pi}).
\label{36a}
\end{equation}
Here $\widetilde{U}^{(0)}(k)=\int \varphi^{(0)}(r) \Phi(r)
\exp(-i{\bf k}{\bf r})d^3r$, where $\varphi^{(0)}(r)$ obeys
Eq.~(\ref{28aa}). Further, substituting $k=\sqrt{n}y$ in the
integral for the condensate depletion $(n-n_0)/n=1/(2\pi)^3
\int_{0}^{+\infty}dk\, 4\pi k^2 n_k/n$, we obtain the familiar
result
\begin{equation}
(n-n_0)/n=8\sqrt{n a^3}/(3\sqrt{\pi})+\cdots,
\label{37}
\end{equation}
$a$ being the scattering length. Inserting (\ref{25}), (\ref{35a}) and
(\ref{36}) into Eq.~(\ref{7}) and using (\ref{36a}), in a
similar manner we derive
\begin{equation}
\varepsilon=\frac{2\pi \hbar^2 a n}{m}
          \Bigl\{1+\frac{128}{15\sqrt{\pi}}\sqrt{na^3}
\Bigl(1-\frac{5}{8}\frac{b}{a}\Bigr)+\cdots\Bigr\},
\label{38}
\end{equation}
where $b \geq 0$ is one more characteristic length defined as
\begin{equation}
b=\frac{1}{4\pi}\int (\nabla\varphi^{(0)}(r))^2\,d^3r.
\label{39}
\end{equation}
As it is seen, the well--known result of papers~\cite{Lee} can be
derived from (\ref{38}) with the choice $b=0$. However, this
approximation is rather crude because the case of the hard--sphere
interaction ($\Phi(r)=0\,(r>a)$ and $\Phi(r)\to\infty\,(r<a)$) is
specified by $b=a$:
\begin{equation}
\varepsilon=\frac{2\pi \hbar^2 a n}{m}
      \Bigl\{1+\frac{16}{5\sqrt{\pi}}\sqrt{na^3}+\cdots\Bigr\}.
\label{40}
\end{equation}
In the general case and, in particular,
for the singular potentials of the Lennard--Jones type we have
$a\not=b$. Remark that the last term in the r.h.s. of (\ref{30})
does not make any contribution into the results given by (\ref{37})
and (\ref{38}). However, the next orders in the expansions of
the energy and depletion depend on its contribution essentially.

Concluding let us take notice of the important points of this Letter
once more. It was demonstrated that thermodynamically consistent
calculations based on (\ref{11}) and (\ref{12}) conclusively result
in Eqs.~(\ref{19})-(\ref{21}).  Therefore, using the Bogoliubov
model with the ``dressed" interaction does not provide the
satisfactory solution of the problem of the strong--coupling Bose
gas. As it was shown, when investigating this subject, one should
go beyond the Bogoliubov scheme. To do this, we developed the
approach reduced to the system of Eqs.~(\ref{27}), (\ref{33}),
(\ref{35a}) and (\ref{36}). This equations reproduce the
well--known result (\ref{37}) for the condensate depletion and
yields the new expansion~(\ref{38}) in powers of $n$ for the energy,
(\ref{40}) being the particular case of the hard spheres.
One can expect alterations for the excitation spectrum, too.

This work was supported by the RFBR Grant No. 97-02-16705.


\end{multicols}

\begin{references}
\bibitem{Bog1} N. N. Bogoliubov, {\it J. Phys. USSR} {\bf 11}, 23
(1947), reprinted in D.~Pines, {\it The many--body problem}
(New York, W.A.  Benjamin, 1961).
\bibitem{Lee} T. D. Lee, K. Huang, C. N. Yang, {\it Phys. Rev.} {\bf
106}, 1135 (1957); K.~A.~Brueckner, K.~Sawada, {\it Phys. Rev.}
{\bf 106}, 1117 (1957); S.~T.~Belyaev, {\it J. Exptl. Theoret. Phys.
(USSR)} {\bf 34}, 433 (1958) [English transl. {\it Sov. Phys. JETP}
{\bf 7}, 299 (1958)]; N.~M.~Hugenholtz, D.~Pines, {\it Phys. Rev.}
{\bf 116}, 489 (1959).
\bibitem{Bog2} N. N. Bogoliubov, {\it Lectures on quantum
statistics}, vol. 1 (Gordon and Breach, New York, 1967), p.39.
\bibitem{Chern1} A. Yu. Cherny, {\it Pair wave functions in a Bose
liquid}, E--print cond-mat/9807120, submitted to Phys. Rev. A.
\bibitem{Chern2} A. Yu. Cherny, A. A. Shanenko,
{\it Phys. Lett. A} {\bf 250}, 170 (1998).
\bibitem{note4} By the leading order we mean zero depletion
of the condensate: $n=n_{0}$.
\bibitem{Shan1} A. A. Shanenko, {\it Phys. Lett. A} {\bf 227},
367 (1997); {\it Phys. Lett. A} {\bf 231}, 414 (1997).
\bibitem{Fetter} A. L. Fetter, J. D. Walecka,
{\it Quantum theory of many--particle systems} (McGraw--Hill,
New York, 1971), p.150.
\bibitem{Bog3} N. N. Bogoliubov, {\it Lectures on quantum
statistics}, vol. 2 (Gordon and Breach, New York, 1970), p.1.
\bibitem{note3} It is not true in the case of complex--value
wave functions. However, in this situation integrands in
(\ref{27}) and (\ref{33}) would include the absolute values
of $\widetilde{\psi}(k)$ and $\widetilde{\psi}_{{\bf p}}({\bf k})$
for which we have $|\widetilde{\psi}_{{\bf p}}({\bf k})|-\sqrt{2}
|\widetilde{\psi}(k)| \propto p^2$ at small $p$ in the case of the
boundary conditions (\ref{4}).
\end{references}
\end{document}